\documentclass{appolb}
\usepackage{epsfig}

\begin{document}
\title{Latest results from Da$\phi$ne%
\thanks{Presented at 'Matter to the deepest', Ustron 2003}%
}
\author{the KLOE collaboration\footnote{A.~Aloisio,
F.~Ambrosino,
A.~Antonelli, 
M.~Antonelli, 
C.~Bacci,
G.~Bencivenni, 
S.~Bertolucci, 
C.~Bini, 
C.~Bloise, 
V.~Bocci
% ...
}\\
Presented by Stefano Di Falco
\address{University and Sezione INFN of Pisa}
}
\maketitle
\begin{abstract}
The Da$\phi$ne Frascati 
$\phi$ factory has continously improved its performances reaching in 2002
an instantaneous luminosity of $8\cdot10^{31}cm^{-2}\ s^{-1}$.
The DEAR experiment, concluded in 2002, has measured the de-excitation of 
kaonic atoms. The KLOE experiment, still running, has measured
several branching ratios for neutral and charged kaons decays,  
$\rho$, $\eta$, $\eta '$, $a_0$ and $f_0$ mesons parameters and,  
via the radiative return, the 
$e^+e^-\rightarrow\pi^+\pi^-$ cross section.  
Preliminary and final results are presented.
  
\end{abstract}
  
\section{Introduction}
The Da$\phi$ne $\phi-factory$ \cite{dafne} is successfully running in Frascati since
1999. It's particular design, with two different rings intersecting only in
the two interaction regions, was intended to reach the extremely 
high luminosity of $5\cdot10^{32}cm^{-2}\ s^{-1}$, allowing 120 bunches per
ring with a lifetime of 2 hours and an average luminosity of 5 $fb^{-1}$
per year. 

During the year 2002, in the KLOE interaction region, a luminosity of 
$8\cdot10^{31}cm^{-2}\ s^{-1}$ has been reached using 45 bunches per ring. 
Beam lifetime has reached the 40 minutes.
The total integrated luminosity, optimized by injecting new bunches during
the data taking,  has
been of 0.3 $fb^{-1}$ in 140 days
corresponding to $\sim2\ pb^{-1}/day$, more than the
double of what had been obtained in 2001.

In the last months of year 2002, 
in the other interaction region, the DEAR experiment
has collected 77 $pb^{-1}$ with a maximum luminosity of
$7\cdot10^{31}cm^{-2}\ s^{-1}$ and 100 bunches circulating in each ring
\cite{drago}.

In the year 2003 major changes have taken place in Da$\phi$ne optics, 
in particular for the KLOE interaction region the layout of the low
beta permanent 
quadrupoles has been substantially modified for easier tuning of the
machine.

The new Da$\phi$ne setup aims to reach a luminosity of 
$2\cdot10^{32}cm^{-2}\ s^{-1}$.

The FINUDA experiment has been installed in the interaction region
previuosly occupied by the DEAR experiment.

In the fall of 2003 KLOE will restart taking data 
with the aim of collecting 2 $fb^{-1}$ before the end of 2004.

The analysis of the data already collected by the DEAR and KLOE experiments 
has produced the results that will be shown in the following.

\section{The DEAR experiment}

\subsection{Physics program}
The DEAR ($Da\phi ne\ Exotic\ Atoms\ Research$) experiment \cite{dear}
has the goal of measuring the isospin dependent 
antikaon-nucleon scattering lengths, $a_0$ and $a_1$, via the
measurement of the $K_\alpha$ line shift ($\epsilon$) and width ($\Gamma$) 
in kaonic hydrogen and deuterium.

From the Deser-Trueman formulas \cite{deser} one has for kaonic hydrogen:

\begin{equation}
 \epsilon +i\Gamma /2 = 412\cdot a_{K^- p}\ eV fm^{-1}
\end{equation}

and for kaonic deuterium:

\begin{equation}
 \epsilon +i\Gamma /2 = 601\cdot a_{K^- d}\ eV fm^{-1}
\end{equation}

and, extracting $a_{K^- n}$ from $a_{K^- d}$, one gets:

\begin{equation}
 a_0=\ 2\cdot a_{K^- p}-a_{K^- n}\ ;\ \ \ a_1=\ a_{K^- n}
\end{equation}

These amplitudes give crucial informations about the meson-nucleon sigma
terms and the strangeness content of the proton.
 
\subsection{Experimental technique and results}

The experimental setup contains a pressurized cryogenic target cell with a
diameter of 11 cm and a thickness of 75 $\mu$m of kapton where the
kaons, produced in the $\phi$ decays, are stopped and captured to form
kaonic atoms. 

The photons produced in the de-excitation of these atoms are detected by 16
CCD-55 ({\em Charged Coupled Devices}).

In a first period of data taking a nitrogen target has been used to tune up
the apparatus: the yield of kaonic nitrogen transitions is indeed 20 times
higher than the kaonic hydrogen ones.

\begin{figure} [thb]
\begin{center}
 \mbox{\epsfig{file=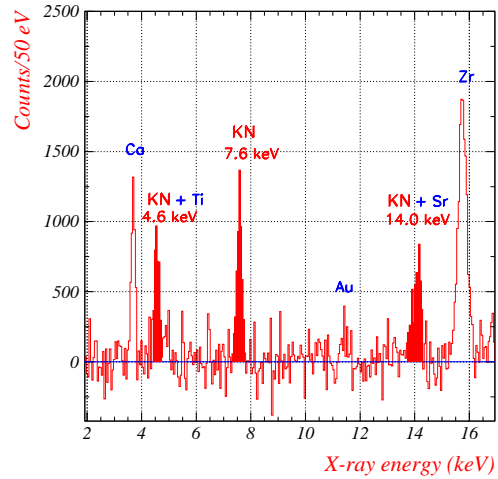,height=7 cm}}
 \caption{\footnotesize Background subtracted energy spectrum obtained with
 KN run of October 2002\cite{catalina}.}
 \label{dearkn}
\end{center}
\end{figure}

Using the $\sim 10\ pb^{-1}$ collected in the october 2002, the yields of
the following 3
transitions have been measured for the first time (fig.\ref{dearkn}),
obtaining the following results\cite{kn}:

\begin{eqnarray}
\nonumber
7->6\ & at\ \ 4.6\ keV:\  2690\pm650\ events & (\ Y=33.7\pm\ 8.1\pm3.4\%\ ) \\
\nonumber
6->5\ & at\ \ 7.6\ keV:\  5320\pm395\ events & (\ Y=55.5\pm\ 4.2\pm5.5\%\ ) \\
\nonumber
5->4\ & at\ 14.0\ keV:\ 1360\pm330\ events & (\ Y=66.4\pm15.6\pm6.4\%\ ) 
\end{eqnarray}

Using the KN data the feasibility of a precision measurement of the charged
kaon mass has also been proved\cite{kmass}.

In december 2002, DEAR has collected $\sim 60\ pb^{-1}$ with an hydrogen
target to measure the $K_\alpha$ line of KH.
In addition, background data with no collision, equivalent to $\sim 30
pb^{-1}$ in collision, have been collected for the study of the background.

Two different analyses based on the global fit of 
the KH spectrum before and after background subtraction
have given as preliminary results for the $K_\alpha$ line shift 
and width\cite{dearkh}:

\begin{equation}
 \epsilon =\ -(162\pm40) \ eV \ ; \ \ \ \Gamma=\ 200\pm80\ eV
\end{equation}

\subsection{Future plans}
An upgrade of the DEAR experiment called SIDDHARTA (SIlicon Drift
Detector for Hadronic Atoms Resarch by Timing Application) is under study.
The new setup will use fast triggerable devices, namely Silicon Drift
Detectors, the trigger being given by the charged kaons entering the cell.
A precision at the eV level is aimed on $K_\alpha$ lines of kaonic hydrogen
and deuterium. The observation of other exotic atoms will also be possible.

\section{The KLOE experiment}

\subsection{Experimental setup}

KLOE (K LOng Experiment) is a general purpose detector \cite{kloeprop}
whose characteristics have been optimized for the study of the CP
violating decays of the neutral kaons.

\begin{figure} [tbh]
\begin{center}
 \mbox{\epsfig{file=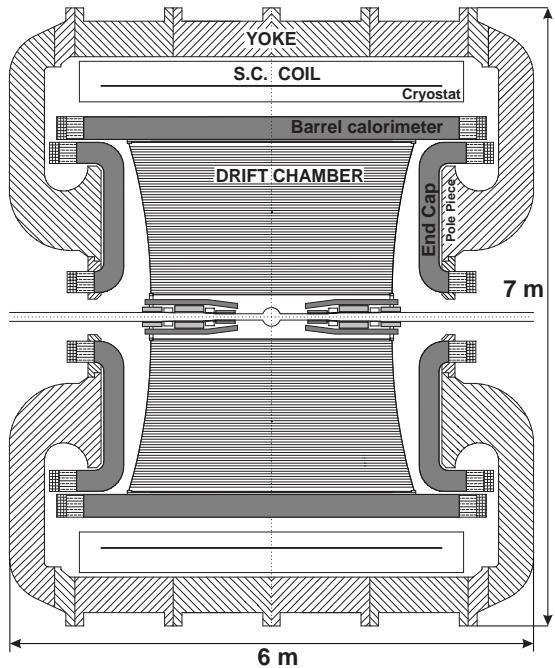,height=9 cm}}
 \caption{\footnotesize Schematic view of KLOE detector.}
 \label{kloedet}
\end{center}
\end{figure}

The main components of the detector are (fig.\ref{kloedet}):
\begin{itemize}
\item a {\em Superconducting Solenoid} giving a magnetic field of 0.52 T.

\item the {\em Drift Chamber} \cite{kloedc}: a cylindrical shape ($\o$=4 m,  
  $l$=3.3 m) filled with a gas mixture, 90\% He+10\% Isobutane,
  instrumented with 12582 stereo sense wires. It gives a momentum resolution   
  $\sigma_p /p\leq 0.4$\% for tracks with $p>100$ MeV and
  $\theta>45^o$. The spatial resolution is 150 $\mu$m in the
  radial direction and 2 mm in the longitudinal one. 

\item the {\em ElectroMagnetic Calorimeter} \cite{kloeemc}: it's a 15$X_0$ 
  sampling calorimeter with lead and scintillating fibers read at both sides by
  photomultipliers. With its cylindrical shape it covers 98\% of the 
  solid angle. The energy resolution is $\sigma_E /E=\
  5.7\%/\sqrt{E(GeV)}$. The time resolution is $\sigma_t =\ 54\
  ps/\sqrt{E(GeV)}\oplus\ 50\ ps$\footnote{Interaction time spread due to
  the bunch length is not included.}. The spatial resolution for completely
  neutral vertices, e.g. $K_L^0 \rightarrow \gamma\gamma$, 
  reconstructed inside the drift chamber is $\sim 1.5$ cm.  

\item the {\em Quadrupole CALorimeter} \cite{kloeqcal}: two smaller sampling 
  calorimeters with lead and scintillating tiles, surrounding the low
  $\beta$ quadrupoles to complete the calorimetric hermeticity.

\end{itemize}

\subsection{KLOE Physics}
Most of the $\phi$ mesons decay in charged kaons (49\%) or neutral
kaons (34\%), so that kaon physics is the largest part of KLOE 
program.

Other particles, produced in $\phi$ decays
with enough abudance to improve the existing
experimental measurements, are  
$\rho$ ($\sim$15\%), $\eta$  ($\sim$1.3\%), $\eta '$  ($O(10^{-4})$), $a_0$
($O(10^{-4})$) and $f_0$ ($O(10^{-4})$).

Furthermore, thanks to the {\em radiative return} that reduces the center
of mass energy of the colliding beams, KLOE can measure 
the hadronic cross section, $e^+ e^-
\rightarrow \pi^+ \pi^- $, in the crucial region 
from the $\pi\pi$ threshold to the $\phi$ resonance.

Results presented in the following are obtained from the analysis of 2000
data (25 $pb^{-1}$) and, in some cases, from the preliminary analysis of 
2001 (170 $pb^{-1}$) and 2002 (280 $pb^{-1}$) data.

\subsubsection{Kaon physics}
Kaons are produced at a $\phi$-factory  in a pure antisymmetric initial 
state with quantum numbers $J^{PC}=1^{--}$. 
At Da$\phi$ne the $\phi$ mesons are produced pratically at
rest, with only a small momentum ($\sim$13 MeV/c) in the plane of the
orbit, so that the kaons are produced practically back to back.

In the case of neutral kaons one has:

\begin{equation}
 |i>\simeq \frac{1}{\sqrt{2}} (|K_L,\vec{p}>|K_S,-\vec{p}>-|K_L,-\vec{p}>|K_S,\vec{p}>)
\end{equation}

This simultaneous production 
allows for a clean tagging of $K_S$($K_L$) looking at the $K_L$($K_S$) on
the other side, having also a precise information on the tagged kaon
momentum.

The two main tagging algorithms in KLOE are:
\begin{itemize}
\item ${\bf K_S}$ {\bf  tagged} by $K_L$ interaction in the calorimeter ($K_L-crash$): 
  $K_L$ interactions are identified using the time of flight and some 
  additional request on the energy deposit\cite{kstag}. The tagging
  efficiency, mostly geometrical, is $\sim 30$\%. The $K_S$ momentum
  is obtained with a resolution of $\sim 2$ MeV and an
  angular resolution of $\sim 1^o$ ($0.3^o$ in $\Phi$).
\item ${\bf K_L}$ {\bf tagged} by $K_S \rightarrow\pi^+ \pi^-$ vertex close to the
  interaction point: $K_S$ decays are identified asking for the $K_S$
  invariant mass and momentum. Tagging efficiency, mainly geometrical, is
  $\sim 70$\%. The resolution on $K_L$ momentum is $\sim 2$ MeV and the
  angular resolution is $\sim 1^o$.
\end{itemize}

Using just the $\sim 2\cdot10^6$ $K_L -crash$ tagged $K_S$ from 17$ pb^{-1}$
collected in 2000, KLOE has measured the ratio of the partial
widths of the main $K_S$ decay modes\cite{kstag}:

\begin{equation} 
  \frac{\Gamma(K_S \rightarrow \pi^+\pi^-(\gamma))}
       {\Gamma(K_S \rightarrow \pi^0\pi^0)}=\ 2.236 \pm 0.003 \pm 0.015
\label{ksratio}
\end{equation}

The measurement is fully inclusive with respect to the radiated photon energy.
The ratio in (\ref{ksratio}) is important not only because it enters in the
{\em double ratio} used to measure $Re(\frac{\epsilon '}{\epsilon})$, but
also because it can be used to derive the strong phases in the amplitudes
$A(K_S\rightarrow\pi\pi)$, investigating  the presence of {\em
  isospin-breaking} electromagnetic phase shifts.

Other interesting $K_S$ branching ratios measured by KLOE are:
\begin{eqnarray}
\nonumber
  BR(K_S\rightarrow \pi^- e^+ \nu)=\ (3.46 \pm 0.09 \pm 0.06)\cdot10^{-4}\\
  BR(K_S\rightarrow \pi^+ e^- \bar{\nu})=\ (3.33 \pm 0.08 \pm 0.05)\cdot10^{-4}
 \label{brksen}
\end{eqnarray}

\noindent the results are obtained from a still preliminary analysis 
of 170 $pb^{-1}$ collected in 2001 and are consistent with the ones
published using the 2000 data sample \cite{ksen}.

Combining the results in (\ref{brksen}) one gets 
the first preliminary measurement of $K_S$ semileptonic asymmetry:
\begin{equation}
 A_S=\ 
 \frac{BR(K_S\rightarrow \pi^- e^+ \nu) - BR(K_S\rightarrow \pi^+ e^-\bar{\nu})}
      {BR(K_S\rightarrow \pi^- e^+ \nu) + BR(K_S\rightarrow \pi^+ e^-\bar{\nu})} =
 (19\pm 17 \pm 6 ) \cdot 10^{-3}
\end{equation}
 
This asymmetry can be compared with the value of the $K_L$ semileptonic
asymmetry \cite{alktev} 
to perform a nice test of CPT conservation.

$K_S$ semileptonic decays can also be used to test the $\Delta S=
\Delta Q$ rule which forbids decays like $K^0\rightarrow \pi^+ e^- \bar{\nu}$. 
The $\Delta S=\Delta Q$ violation is commonly expressed via:
\begin{equation} 
 Re\ x_+ =\ \frac{1}{2}
 \frac{BR_S(\pi e \nu)/ \tau_S - BR_L(\pi e \nu)/ \tau_L}
 {BR_S(\pi e \nu)/ \tau_S + BR_L(\pi e \nu)/ \tau_L}
\label{xplusdef}
\end{equation}

If CPT holds, $\Delta S=\Delta Q$ implies $x_+ = 0$. The standard model
allows $\Delta S=\Delta Q$ violation only at next to leading order 
predicting $Re\ x_+\sim 10^{-7}$. KLOE preliminary result (2001 data sample)
is:
\begin{equation} 
 Re\ x_+ =\ (2.2 \pm 5.3 \pm 3.5)\cdot 10^{-3}
\label{xplus}
\end{equation}

\noindent in agreement with the comparabale result obtained by CPLEAR 
\cite{cplear}.

Using an energy scan around the
$\phi$ peak, KLOE has also measured the $K_S$ mass\cite{ksmass}:
\begin{equation}
  m(K_S)=\ 497.583 \pm 0.005 \pm 0.020\ MeV
\end{equation}
\noindent by normalizing the energy scale to the precise determination of 
the $\phi$ mass obtained by resonant beam depolarisation at Novosibirsk\cite{cmd2scan}.

$K_L$ neutral decays are reconstructed using the 
momentum information obtained from the $K_S\rightarrow\pi^+ \pi^-$ vertex
and the time and position of the energy deposited in the calorimeter by the
photons produced in the decays. These informations are sufficient to
 completely reconstruct neutral vertices like $K_L\rightarrow\pi^0\pi^0$ or 
$K_L\rightarrow\gamma\gamma$ with a spatial resolution of $\sim 1.5$ cm.

Exploiting the simple dynamics of the $\gamma\gamma$ decay it's possible 
to obtain a precise measurement of the corresponding branching 
ratio\cite{klgg} (362 $pb^{-1}$ analyzed):

\begin{equation}
  BR(K_L\rightarrow \gamma\gamma) = (2.793 \pm 0.022 \pm 0.024)\cdot 10^{-3} 
\end{equation}

\noindent in agreement with the recent  NA48 result\cite{klggna48}.

Preliminary results on $K_L$  charged decays are also in
good agreement with the PDG values\cite{klch}. 
In particular, a final error much better than 1\% is expected on $K_{l3}$
decays, improving the error on the kaon form factors ($\lambda_+$, 
$\lambda_0$), the $K_L$ lifetime ($\tau_L$) and, finally, on the
CKM matrix element $V_{us}$.

Charged kaons are tagged using the two body decays into $\mu^\pm\nu$ and
 $\pi^\pm\pi^0$. Again one kaon is tagged looking at the other 
kaon on the opposite side. 

Using the momentum and time of flight
informations obtained by the drift chamber and the calorimeter it's
possible to isolate the different decay channels\cite{kppp}.

In particular for the $\pi^\pm \pi^0\pi^0$ channel one gets:
\begin{equation}
 BR(K^\pm\rightarrow\pi^\pm \pi^0\pi^0) =\ 1.807\pm 0.008\pm 0.018\% \ \ \
 (112\ pb^{-1})
\end{equation}

Looking at the charge asymmetries of the slopes of the Dalitz plot for this
channel it's possible to have an evidence of 
direct CP violation in charged kaon decays. 
A preliminary fit of the Dalitz plot 
with just $6.3\ pb^{-1}$ has obtained\footnote{Errors are
  statistical only.}: 
\begin{equation}
g=0.607\pm0.026;\ \ \
h=0.026\pm0.027;\ \ \
k=0.0080\pm0.0037
\end{equation}

\subsubsection{Non-kaon physics at $\phi$ resonance}

From the fit of the Dalitz plot for 
the $\phi\rightarrow\pi^+\pi^-\pi^0$ decays, it's
possible to obtain the $\rho$-meson parameters for its three charge states
and the cross section for $\e^+e^-\rightarrow \omega\pi^0$
with $\omega\rightarrow\pi^+\pi^-$. 
The result obtained with 17 $pb^{-1}$ from 2000 data sample is\cite{rhopi}:
\begin{eqnarray}
\nonumber
m_\rho =\ 775\pm 0.5\pm 0.3 \ MeV;  \ \ \ \ \ \ \ \ \
\Gamma_\rho =\ 143.9\pm1.3\pm1.1\ MeV
\\
\nonumber
m_0-m_\pm=\ 0.4\pm0.7\pm0.6\ MeV; \ \ m_+-m_-=\ 1.5\pm0.8\pm0.7\ MeV\\
\nonumber
\sigma(e^+ e^-\rightarrow\omega\pi^0\rightarrow\pi^+\pi^-\pi^0)= 92\pm15\
pb \ \ \ \ \ \ \ \ \ \ \ \ \
\end{eqnarray}

Using 2000 data KLOE has also measured the relative branching ratio for 
$\phi$ decays in $\eta '\gamma$ and $\eta \gamma$\cite{etaetap}:
\begin{equation}
 \frac{BR(\phi\rightarrow\eta '\gamma)}
      {BR(\phi\rightarrow\eta \gamma)}=
      (4.70\pm 0.47\pm0.31)\cdot10^{-3}
\end{equation}
\noindent from where one can extract the pseudoscalar mixing angle
($\theta_P$) and the gluon content in $\eta '$ ($Z'$):
\begin{equation}
\theta_P=\ (12.9 ^{+1.9}_{-1.6}\ )^o\ ;\ \ \ (Z')^2=\ 0.06^{+0.09}_{-0.06}
\end{equation}

Scalar mesons are detected at KLOE through their decay channels:
\mbox{$a_0\rightarrow\eta\pi^0\gamma$} \cite{a0} and
$f_0\rightarrow\pi^0\pi^0\gamma$ \cite{f0}. From the integration of
the $\eta\pi^0$ and $\pi^0\pi^0$ mass spectra of 2001 and 2002, these 
preliminary updated branching ratios are found:
\begin{eqnarray}
 \nonumber
 BR(\phi\rightarrow \eta\pi^0\gamma)=\ (0.85\pm0.05\pm0.06)\cdot 10^{-4}\\
 \nonumber
 BR(\phi\rightarrow \pi^0\pi^0\gamma)=\ (1.09\pm0.03\pm0.05)\cdot 10^{-4}
\end{eqnarray}
A fit of the Dalitz plots to extract the $a_0$ and $f_0$ contributions is in progress.
 
\subsubsection{Measurement of hadronic cross section}
The determination of hadronic cross section
$\sigma(e^+e^-\rightarrow\pi^+\pi^-)$ in the energy region from $\pi\pi$
threshold to the $\phi$ resonance is still 
one of the main source of error for the
theoretical estimate of the muon anomaly ($a_\mu$) \cite{jeger}.

Furthermore, recent results obtained in this region from $e^+e^-$
and $\tau$ data show an inconsistency of some percent \cite{davier}.

Despite of the fact that Da$\phi$ne runs at fixed energy, 
$\sqrt{s}=m_\phi$, KLOE can measure the hadronic cross section via the
{\em radiative return}\cite{kuhn}: Initial State Radiation (ISR) from the
beams can lower the value of $\sqrt{s}$ down to the $\pi^+\pi^-$ threshold.

KLOE $\pi\pi\gamma$ events are selected asking for two charged tracks
coming 
from the interaction region with an angle respect to the beam pipe
$50^o<\theta<130^o$. Additional cuts on the momentum, $p_T>160$
MeV or $|p_Z|> 90$ MeV, are used to reject tracks spiraling along the beam
line. 

In order to suppress the events in which the $\pi\pi$ invariant mass is
reduced by the presence of a Final State Radiated (FSR) photon, a cut on the
missing momentum angle, $\theta_{miss}<15^o\ or\ \theta_{miss}>165^o$, is
imposed.

Background from Bhabha scattering, $\phi\rightarrow\pi^+\pi^-\pi^0$
decays and $e^+e^-\rightarrow\mu^+\mu^-\gamma$ events is suppressed using a
particle identification method based on the time of flight, the energy
deposit in the calorimeter and the kinematic closure of the event in the
hypothesis of only one photon in the final state\cite{eps}.

Residual background is evaluated from the fit of the mass distribution of
the charged particles and subtracted.

The yield of the selection is $\sim 11000\ events/pb^{-1}$.

The cross section is obtained by dividing for the luminosity measured with
large angle Bhabha events using the BABAYAGA generator\cite{babayaga}.

\begin{figure} [thb]
\begin{center}
 \mbox{\epsfig{file=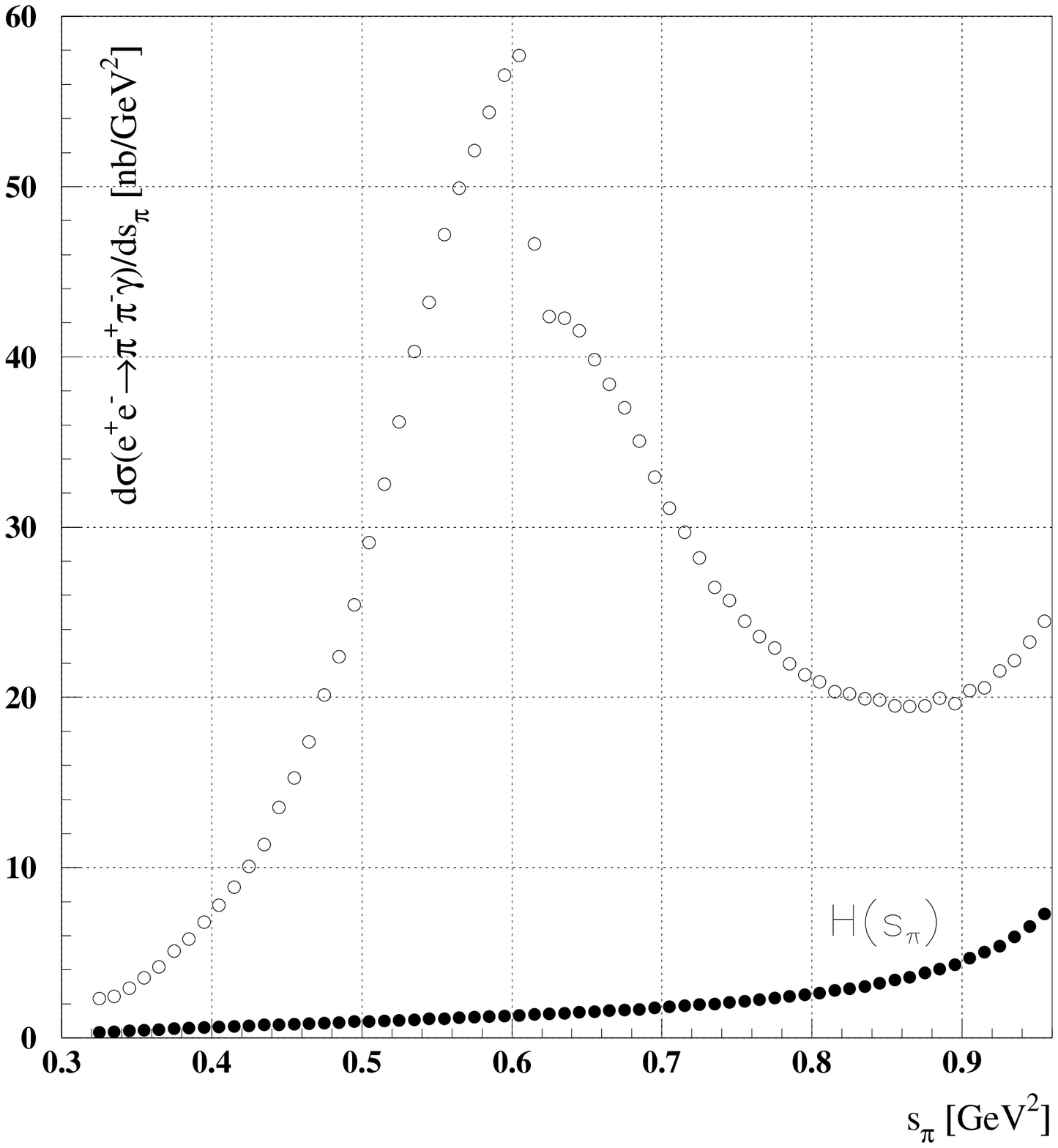,height=6.2 cm}}
 \mbox{\epsfig{file=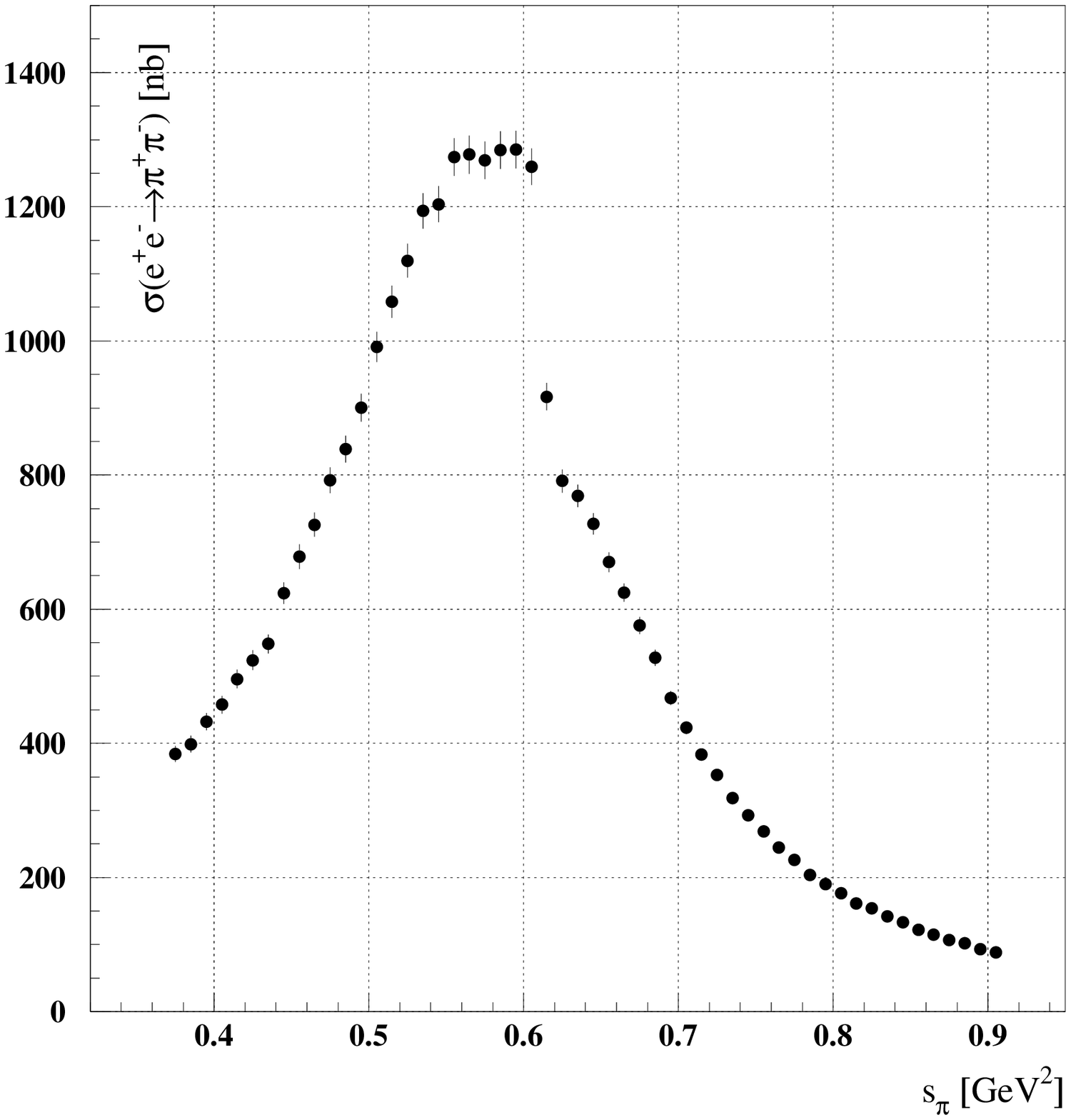,height=6.2 cm}}
 \caption{\footnotesize Left: Cross section for
 $e^+e^-\rightarrow\pi^+\pi^-\gamma$. The radiator function $H$ is also
 shown. Right: Bare cross section for $e^+e^-\rightarrow\pi^+\pi^-$.}
 \label{shad}
\end{center}
\end{figure}

The $e^+e^-\rightarrow\pi^+\pi^-$ cross section 
is obtained from the 
$e^+e^-\rightarrow\pi^+\pi^-(\gamma)$ cross section by dividing for the 
{\em radiation  function} (fig.\ref{shad}.a) given by the PHOKHARA event 
generator\cite{phokhara} used with $F_\pi=1$ and vacuum polarization 
off\footnote{The method is exact only if the contribution of FSR is
  negligible. This holds only at leading order (see \cite{phokhara})
and the effect of residual FSR will be quoted as a 1\% 
in the final systematical error.} and multiplying by the appropriate
kinematic factor.

After correcting for the vacuum polarization the bare
$e^+e^-\rightarrow\pi^+\pi^-$ cross section shown in fig.\ref{shad}.b is
obtained.

The corresponding preliminary value for the $\pi\pi$ contribution to
$a_\mu$ in the
interval $0.37<s<0.93\ GeV^2$   is:
\begin{equation}
 a_\mu^{\pi\pi}= \ 378.4\pm0.8_{stat}\pm4.5_{syst}\pm3.0_{theo}\pm3.8_{FSR}
\label{amu}
\end{equation}
\noindent where the theoretical error comes from the knowledge of the 
radiation function,
of the vacuum polarization and of the Bhabha cross section used to determine the
luminosity.

The result (\ref{amu}) is in good agreement with the one obtained by
CMD2 with the energy scan method\cite{cmd2} and confirms a discrepancy with
 $\tau$ data.

\section*{Aknowledgments}
It's a pleasure to thank C. Curceanu (Petrascu) for the help she gave in
presenting the DEAR results and all the Da$\phi$ne team for the nice
successful work
they did to improve the accelerator performances.

\end{document}